\documentclass[12pt]{iopart}
\usepackage[dvipdfm]{graphicx}

\usepackage{iopams}
\usepackage{setstack}
\usepackage{amsthm}

\newtheorem{thm}{Theorem}

\newtheorem{lem}{Lemma}
\newtheorem{cor}{Corollary}

\newtheorem*{ass}{Assumption}

\begin{document}
\title{Loop series expansion with propagation diagrams}
\author{Yusuke Watanabe and Kenji Fukumizu}
\address{Institute of Statistical Mathematics, 4-6-7,
Minami-Azabu, Minato-Ku, Tokyo, 106-8569}
\eads{\mailto{watay@ism.ac.jp}}

\begin{abstract}
The Bethe approximation is a successful method for approximating
partition functions of probabilistic models associated with a graph.
Recently, Chertkov and Chernyak derived an interesting formula
called ``Loop Series Expansion'',
which is an expansion of the partition function.
The main term of the series is the Bethe approximation while other terms are 
labelled by subgraphs called generalized loops.
In this paper, we derive a loop series expansion of binary pairwise Markov random fields
with ``propagation diagrams'', which describe rules how ``first messages''
 and ``secondary messages'' propagate.
Our approach allows to express the loop series in the form of a polynomial with
coefficients positive
integers.
Using the propagation diagrams, we establish a new formula that shows a
 relation between the exact marginal probabilities 
and their Bethe approximations.
\end{abstract}
\submitto{\JPA}

\section{Introduction}　
A Markov random field (MRF) associated with a graph is given by a joint
probability distribution over a set of variables.
In the associated graph,
the nodes represent variables and the edges represent probabilistic
dependence between variables. 
A typical example of a MRF is a Gibbs distribution of the Ising model on a
finite lattice.
The joint distribution is often given in an unnormalized form,
and the normalization factor of a MRF is called a partition function.

The main topic of this paper is computation of the partition function
and the marginal distributions of a MRF with discrete variables. 
This problem is in general computationally
intractable for a large number of variables, and some approximation method is required.
Among many approximation methods, the Bethe approximation \cite{Bethe}
has attracted renewed interest of computer scientists;
it is equivalent to Loopy Belief Propagation
(LBP) algorithm \cite{empiricalstudy,GBP}, which has been successfully used for 
many applications such as error correcting codes, 
inference on graphs, image processing, and so on
\cite{LDPC,Turbo,LowLevelVision}.

Chertkov and Chernyak \cite{LoopPRE,Loop} give a new and
interesting formula called loop series expansion, which expresses the partition
function in terms of a finite series.
The first term is the Bethe approximation, and the others are labelled by
so-called generalized loops.
The Bethe approximation can be corrected with this formula, though summing
up all the terms requires computational efforts exponential to the
number of linearly independent cycles.

In this paper we propose an alternative diagram-based method for deriving the loop series
expansion formula. In our approach, we define secondary messages, which
are orthogonal to the messages used in the LBP algorithm, and show that
they satisfy a set of rules as they propagate.
For each node and edge, we associate parameters
$\{\gamma_i\}$ and $\{\beta_{ij}\}$, respectively;
$\gamma_i$ is related to the approximated marginal of a node $i$, and
$\beta_{ij}$ to the approximated correlation of adjacent nodes $i$
and $j$. 
The loop series is represented by a polynomial of these variables 
with coefficients positive integers.
This positivity is useful for deriving a bound on the number of
generalized loops.

The main result of this paper is theorem \ref{thmMarginalExpansion},
with which we can calculate the true marginal probabilities in terms
of the beliefs at the convergence of the LBP, $\{\gamma_i\}$, and $\{\beta_{ij}\}$.
The terms in the formula of the marginals depend on the topological 
structure of the graph.

This paper is organized as follows.
In section 2, we briefly review the definition of pairwise MRF,
the Bethe approximation and the LBP algorithm.
In section 3, we characterize the Bethe approximation as the fixed points of
the LBP and deduce the fixed point equation in theorem
\ref{BPeqThm}. In section 4, we define first and secondary messages, and
study their propagation rules. These rules are fundamental tools for our
analysis.
In section 5, we derive the loop series formula, and compare it with the
results of Chertkov and Chernyak. 
We deduce consequences of our representation of the expansion:
the connection to the partition function of the Ising model 
(with uniform coupling constant and external field) and 
the upper bound on the number of generalized loops.
In section 6, we prove an expansion formula for the true marginal
probability, and provide some examples.

\section{Bethe approximation and loopy belief propagation algorithm}
\subsection{Pairwise Markov random field}
We introduce a probabilistic model considered in this paper,
MRF of binary states with pairwise interactions.
Let $G:=(V,E)$ be a connected undirected graph, where
$V=\{1,\ldots,N \}$ is a set of nodes and 
$E\subset \{(i,j); 1\leq i < j \leq N \}$
is a set of undirected edges.
Each node $i \in V$ is associated with a binary space $\chi_i=\{\pm 1\}$.
We make a set of directed edges from $E$ by $\vec{E}=\{(i,j),(j,i);(i,j)\in E
\}$.
The neighbours of $i$ is denoted by
$N(i) \subset V$, and $d_i= |N(i)|$ is called the degree of $i$.
A joint probability distribution on the graph $G$ is given by the form:
\begin{eqnarray}
p(x)=\frac{1}{Z} \prod_{ij\in E} \psi_{ij}(x_i,x_j) 
\prod_{i \in V} \phi_{i}(x_i), \label{MRF}
\end{eqnarray}
where $\psi_{ij}(x_i,x_j):\chi_i \times \chi_j \rightarrow
\mathbb{R}_{> 0}$ and  
$\phi_i : \chi_i \rightarrow \mathbb{R}_{> 0}$ are positive
functions called compatibility functions.
The normalization factor $Z$ is called the partition function.
A set of random variables which has a probability distribution in the
form of (\ref{MRF}) is called a Markov random
field (MRF) or an undirected graphical model on the graph $G$.
This class of probability distributions is equivalent to the Ising model with
arbitrary coupling constants and local magnetic fields.
In traditional literatures of statistical physics, a graph $G$ is often
given by an infinite lattice, but as per recent interest,
especially in computer science, $G$ has an arbitrary topology with finite nodes.

Without loss of generality, univariate compatibility functions $ \phi_{i}$ can 
be neglected because they can be included in bivariate compatibility functions $\psi_{ij}$.
This operation does not affect the Bethe approximation and the LBP algorithm
given below; we assume it as per the following.

\subsection{Loopy belief propagation algorithm}
The LBP algorithm computes the Bethe approximation of
the  partition function and the marginal distribution of each node
with the message passing method \cite{Pearl,empiricalstudy,GBP}.
This algorithm is summarized as follows.
\begin{enumerate}
\item Initialization: \\
For all $(j,i) \in \vec{E}$, the message from $i$ to $j$ is a vector
 $m_{(j,i)}^{0} \in \large{\mathbb{R}}^{2} $. Initialize as
\begin{eqnarray}
m_{(j,i)}^{0}(x_j)=1  \quad \forall x_j \in \chi_j .
\end{eqnarray}

\item Message Passing: \label{MessagePassing}\\
For each $t=0,1,\ldots$, update the messages by
\begin{equation}
m^{t+1}_{(j,i)}(x_j)=\omega \sum_{x_i \in \chi_i} \psi_{ji}(x_j,x_i)
\prod_{k \in N(i) \backslash  \{j\} } 
m^t_{(i,k)}(x_i), \label{BPupdate}
\end{equation}
until it converges. Finally we obtain
$\{m^{*}_{(j,i)}\}_{(j,i)\in \vec{E}}$.

\item 
Approximated marginals and the partition function are computed
by the following formulas:
\begin{eqnarray}
\fl
b_i(x_i):= \omega \prod_{j \in N(i)}m^{*}_{(i,j)}(x_i), \label{1marginal}\\
\fl
b_{ji}(x_j,x_i):= \omega \quad \psi_{ji}(x_j,x_i)  \prod_{k \in
 N(j)\backslash \{i\}} m^*_{(j,k)}(x_j) 
\prod_{k' \in N(i) \backslash \{j\}} m^*_{(i,k')}(x_i) ,\label{2marginal}
\\
\fl
\log Z_B := \sum_{(j,i) \in {E}}\sum_{x_j, x_i}
 b_{ji}(x_j,x_i)\log\psi_{ji}(x_j,x_i)   
 - \sum_{(j,i) \in {E}}\sum_{x_j x_i} b_{ji}(x_j,x_i)\log
 b_{ji}(x_j,x_i) \nonumber  \\
+ \sum_{i \in V}(d_i-1)\sum_{x_i}b_i(x_i)\log b_i(x_i), \label{Bethe}
\end{eqnarray}
where $\omega$ are appropriate normalization constants,
$b_i$ are called beliefs, and $Z_B$ is called the Bethe approximation of the partition function.
\end{enumerate}
In step (\ref{MessagePassing}), there is ambiguity as to the order of
updating the messages. 
We do not specify the order, because the fixed points of LBP algorithm
do not depend on its choice.
Note that this LBP algorithm does not necessarily converge, and there
may be more than one fixed points unless the interactions are sufficiently
weak \cite{Gibbsmeasure}. 

\section{Fixed point equation of LBP} 
When LBP converges, any converged messages $\{m^{*}_{(j,i)}\}_{(j,i)\in \vec{E}}$
satisfy a certain equation shown in theorem \ref{BPeqThm}. 
By this theorem we show that the converged messages can be normalized
simultaneously and we define 
$\{\mu_{(j,i)}\}_{(j,i)_{\in\vec{E}}}$ called first messages.

\subsection{Graph operations} \label{3.1}
\begin{figure}
\begin{minipage}{.4\linewidth}
\vspace{0.6cm}
\hspace{2.0cm}
\includegraphics[scale=0.4]{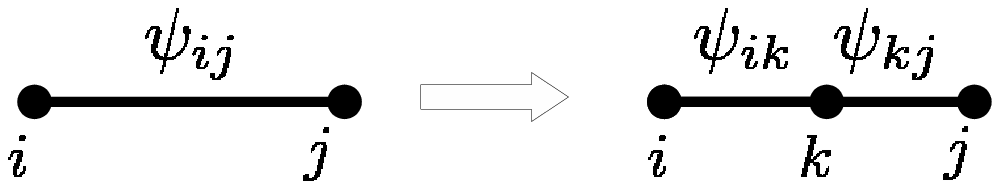}
\caption{A new node can be added on an edge.}
\label{fig1}
\end{minipage}
\hspace{-0.1cm}
\begin{minipage}{.55\linewidth}
\hspace{2.0cm}
\includegraphics[scale=0.5]{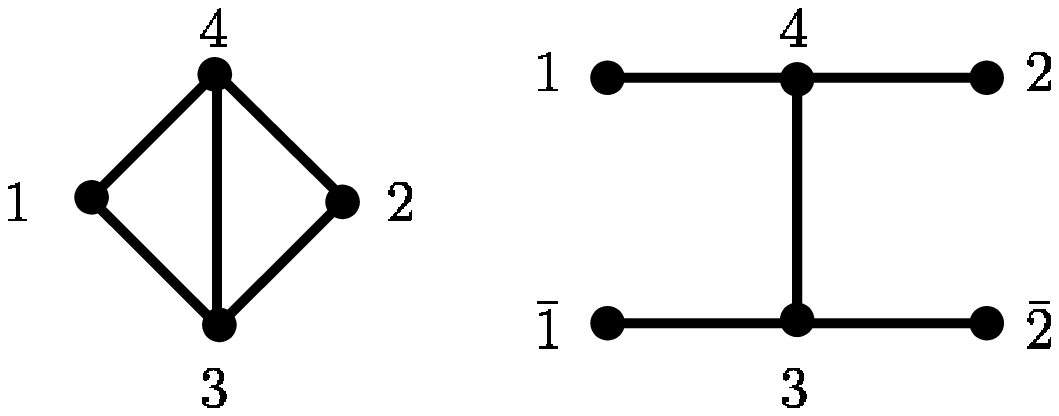} 
\caption{An example of $G$ and $\hat{G}$.}
\label{fig2}
\end{minipage}
\end{figure}

First, we remark that we can always
add a new node
without changing the marginals and beliefs of the others. 
For an edge $(i,j)$, we can add a node $k$ between $i$ and $j$ as in
\fref{fig1}
with new compatibility functions $\psi_{ik},\psi_{kj}$ satisfying
\begin{eqnarray}
 \psi_{ij}(x_i,x_j) = \sum_{x_k} \psi_{ik}(x_i,x_k)\psi_{kj}(x_k,x_j).
\end{eqnarray}
 This operation will be used
implicitly many times in this paper.
Adding new nodes, if necessary, we can always assume that ``there are sufficiently
many nodes of degree two''. 
%

Next we define a graph $\hat{G}$ by $G$.
Let $L:=|E|-|V|+1$, the number of linearly independent cycles. 
Cutting and duplicating $L$ nodes of degree two appropriately, we obtain a connected tree
$\hat{G}$,
since we assume that there are sufficiently many nodes of
degree two \cite{Yellen}. See \fref{fig2}. 
Renumbering the nodes of $V$, we assume that
the cut nodes are numbered by $\{1,\ldots,L\}$.
We define $\hat{G}=(\hat{V},\hat{E})$ by
$\hat{V_1}=\{1,\ldots,L\} \cup  \{\bar{1},\ldots,\bar{L}\}$,
$\hat{V_2}=\{L+1,\ldots,N \}$ and 
$\hat{V}=\hat{V_2} \cup \hat{V_1}$.
$\hat{E}$ is also naturally defined.
We call $\hat{V}_1$ leaf nodes.
\subsection{Belief propagation equations}
Using the converged messages $\{m^{*}_{(j,i)}\}$,
we define messages coming into the leaf nodes of the graph $\hat{G}$ as follows.
Let $s$ be a cut node with the neighbour $N(s)=\{u,v\}$ in $G$, and
let $(s,v),(\bar{s},u) \in \hat{E}$
be the edges at the duplicated nodes 
We define $\mu_{{s}} \propto m^{*}_{(s,u)}$,
$\mu_{\bar{s}} \propto m^{*}_{(s,v)}$ and normalize them by
$\sum_{x_s} \mu_s(x_s) \mu_{\bar{s}}(x_s)=1$.
Generalizations of the transfer matrices are defined by
\begin{eqnarray}
T^{x_1,\ldots,x_L}_{\quad x_{\bar{1}},\ldots,x_{\bar{L}}} = 
\sum_{x_{L+1},\ldots,x_{N}} \prod_{(i,j) \in \hat{E}}
\psi_{ij}(x_i,x_j), \label{t}
\end{eqnarray}
\vspace{-0.5cm}
\begin{eqnarray}
 {T_s}^{x_s}_{\quad {x_{\bar{s}}}}=
\sum_{x_1,\ldots,x_{s-1},\atop x_{s+1},\ldots,x_L}
\sum_{x_{\bar{1}},\ldots,x_{\bar{s-1}},\atop
x_{\bar{s+1}},\ldots,x_{\bar{L}} }
T^{x_1,\ldots,x_L}_{\quad x_{\bar{1}},\ldots,x_{\bar{L}}}
\prod^{L}_{m=1 \atop m \neq s}\mu_m(x_m)\mu_{\bar{m}}(x_{\bar{m}}) .\label{ts}
\end{eqnarray}
Since $\{m^{*}_{(j,i)}\}$ is the convergence point of the LBP algorithm, 
it is easy to see that there is $\alpha_s > 0$ that satisfies
$\alpha_s \mu_s(x_s)=  \sum_{{x_s}'} {T_s}^{{x_s}'}_{\quad {x_s}}
 \mu_s({x_s}') .$
The following theorem states that all $\alpha_s$ are equal to $Z_B$. 
%
%
\begin{thm}\label{BPeqThm}
for $s=1,\ldots,L$
\begin{eqnarray}
\label{BPeq1}
Z_B \mu_s(x_{\bar{s}})=  \sum_{{x_s}} {T_s}^{{x_s}}_{\quad {x_{\bar{s}}}} \mu_s({x_s}) 
\quad & x_{\bar{s}} \in \chi_{\bar{s}} ,\\
\label{BPeq2}
Z_B \mu_{\bar{s}}(x_s)= \sum_{{x_{\bar{s}}}} {T_s}^{{x_s}}_{\quad {x_{\bar{s}}}}
 \mu_{\bar{s}}({x_{\bar{s}}})
\quad & {x_s} \in \chi_{{s}}.
\end{eqnarray}
\end{thm}
\begin{proof}
From
$ \alpha_s \mu_s(x_s)= \sum_{{x_s}'} {T_s}^{{x_s}'}_{\quad {x_s}} \mu_s({x_s}') $,
$\alpha_s$ and $\alpha_{\bar{s}}$ are the Perron-Frobenius eigenvalues of the matrix
${T_s}^{{x_s}'}_{\quad{x_s}}$ and its transpose, respectively.
Therefore $\alpha_s = \alpha_{\bar{s}}$ for $1 \leq s \leq L$.
Next, we prove $\alpha_s = \alpha_l$ for $1 \leq s,l \leq L$.
As we normalize $\sum_{x_s} \mu_s(x_s) \mu_{\bar{s}}(x_s)=1$,
\begin{eqnarray}
\alpha_{s}^{}& = \sum_{x_s} \{\alpha_{s}^{} \mu_s(x_s)\}
 \mu_{\bar{s}}(x_s) \nonumber \\
& =\sum_{x_1\ldots,x_L}\sum_{x_{\bar{1}}\ldots,x_{\bar{L}}}
T_{\quad x_{\bar{1}},\ldots,x_{\bar{L}}}^{x_1,\ldots,x_L} 
\prod^{L}_{m=1}\mu_m(x_m)\mu_{\bar{m}}(x_{\bar{m}}) \nonumber \\
& =\alpha_{l}^{}, \nonumber
\end{eqnarray}
which shows $\alpha_l = \alpha_s = \alpha$.
Finally we prove $\alpha=Z_B$.
By dividing some $\psi$ function by $\alpha$ from the first,
we can assume that $\alpha=1$.
Then it is sufficient to prove 
$\log Z_B =0$.
By distributing the messages $\{\mu_s\}_ {s\in\hat{V}_1}$ on the tree
 $\hat{G}$ from its leaf nodes $\hat{V}_1$  without normalization, we
 define a message $\mu_{(j,i)}$ at each edge $(j,i)\in \vec{E}$.
Because $\hat{G}$ is a tree, the messages 
$\{\mu_{(j,i)}\}_{(j,i)\in\vec{E}}$ are uniquely defined step by step
 from the leaf nodes.
By the assumption $\alpha=1$, we have $\mu_s=\mu_{(s,u)}$.
Therefore,
\begin{eqnarray}
 \mu_{(j,i)}(x_j)= \sum_{x_i \in \chi_i} \psi_{ji}(x_j,x_i)
\prod_{s \in N(i) \backslash  \{j\} } 
\mu_{(i,s)}(x_i),  \qquad  {}^{\forall}(j,i) \in \vec{E}. \label{BPupdatec}
\end{eqnarray}
By the relation
\begin{equation}
\fl 
\sum_{x_i} \prod_{j \in N(i)}\mu_{(i,j)}(x_i) =
\sum_{x_1\ldots,x_L}\sum_{x_1\ldots,x_L} 
T_{\quad x_{\bar{1}},\ldots,x_{\bar{L}}}^{ x_1,\ldots,x_L} 
\prod^{L}_{m=1}\mu_m(x_m)\mu_{\bar{m}}(x_{\bar{m}}) = 1, \label{equal1}
\end{equation}
we obtain
\begin{eqnarray}
b_i(x_i)=  \prod_{j \in N(i)}\mu_{(i,j)}(x_i) \label{1marginalc},\\
b_{ji}(x_j,x_i)=  \psi_{ji}(x_j,x_i)  \prod_{s \in
 N(j)\backslash \{i\}} \mu_{(j,s)}(x_j) 
\prod_{s' \in N(i) \backslash \{j\}} \mu_{(i,s')}(x_i).
\label{2marginalc}
\end{eqnarray}
The assertion follows from 
putting (\ref{1marginalc}) and (\ref{2marginalc}) into the definition of
 $Z_B$ (\ref{Bethe}).
\end{proof}
As used in the proof of above theorem, the normalization $Z_B=1$ is convenient;
in the rest of this paper we assume the following.
\begin{ass}
 By normalizing one of $\{\psi_{ij}\}$, we assume $Z_B=1$.
\end{ass}

In the proof of the above theorem, we defined
the messages $\mu_{(j,i)} \propto m^{*}_{(j,i)}$
satisfying (\ref{BPupdatec}),(\ref{1marginalc}) and (\ref{2marginalc}).
We call $\mu_{(j,i)}$ (normalized) first messages.
While these conditions are similar to
(\ref{BPupdate}),(\ref{1marginal}) and (\ref{2marginal}),
an important difference is disappearance of the normalization
constants $\omega$.
By the above assumption, the first messages satisfy \eref{BPupdatec}.
Notice that we define $\{\mu_{(j,i)}\}_{(j,i)\in \vec{E}}$ on
the graph $G$, not only on the graph $\hat{G}$.

This equation is a generalization of recursive expression
for calculating free energy of the Bethe lattice \cite{ExactlySolvedModels}.
For each fixed point of the LBP algorithm $\{m^{*}_{(j,i)}\}$, there is a
solution of equation (\ref{BPeq1}),(\ref{BPeq2}). On the other hand, for each solution of 
(\ref{BPeq1}) and (\ref{BPeq2}), there is a LBP fixed point.

\section{Propagation diagrams}
%
We proved in the previous section,
if we normalize $\sum_{x_s} \mu_s(x_s) \mu_{\bar{s}}(x_s)=1$,
the Bethe-approximated partition function is given by
\begin{eqnarray}
Z_B =
\sum_{x_{{1}}\ldots,x_{{L}}}
\sum_{x_{\bar{1}}\ldots,x_{\bar{L}}}
 T_{\quad x_{\bar{1}},\ldots,x_{\bar{L}}}^{x_1,\ldots,x_L} 
\prod^{L}_{m=1}\mu_m(x_m)\mu_{\bar{m}}(x_{\bar{m}}), \label{BetheZ}
\end{eqnarray}
while the true partition function is 
\begin{eqnarray}
 Z=
\sum_{x_{{1}}\ldots,x_{{L}}}
 T_{\quad x_{{1}},\ldots,x_{{L}}}^{x_1,\ldots,x_L} 
=\sum_{x_{{1}}\ldots,x_{{L}}}
\sum_{x_{\bar{1}}\ldots,x_{\bar{L}}}
 T_{\quad x_{\bar{1}},\ldots,x_{\bar{L}}}^{x_1,\ldots,x_L} 
\prod^{L}_{m=1} \delta_{x_{m},x_{\bar{m}}}. \label{trueZ}
\end{eqnarray}
Let us define vectors $\nu_{s}$ and  $\nu_{\bar{s}}$ for $s=1, \ldots, L$ as to satisfy
\begin{eqnarray}
\fl
 \sum_{x_s}\mu_{s}(x_s)\nu_{\bar{s}}(x_s)=0 ,\quad
 \sum_{x_s}\mu_{\bar{s}}(x_s)\nu_{{s}}(x_s)=0, \quad
\sum_{x_s} \nu_s(x_s) \nu_{\bar{s}}(x_s)=1.
\end{eqnarray}
Then, we have a decomposition of the unit matrix
\begin{eqnarray}
 \delta_{x_{s},x_{\bar{s}}}=
  \mu_{s}(x_s)\mu_{\bar{s}}(x_{\bar{s}})+\nu_{s}(x_s)\nu_{\bar{s}}(x_{\bar{s}}). \label{deltaidentity}
\end{eqnarray}
We can expand \eref{trueZ} using \eref{deltaidentity} in a sum of $2^{L}$
terms. The first term is obviously the Bethe approximated partition function
\eref{BetheZ}.
But the explicit form of the remaining $2^{L}-1$ terms is not obvious.
In this section we define secondary messages $\{\nu(i,j)\}_{(i,j)\in\vec{E}}$ and derive rules, which
describe how these messages propagate, for deriving the remaining terms.
\subsection{Definition of secondary messages}\label{4.1}
\begin{figure}[b]
\begin{center}
\includegraphics[scale=0.4]{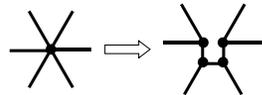}  
\caption{A node of degree $n$ can be split into $n-2$ nodes of degree
  3.}
 \label{fig3}
\end{center}
  \end{figure}

\begin{figure}
\hspace{-0.9cm}
\begin{minipage}{.5\linewidth}
\vspace{0.25cm}
\hspace{3.0cm}
 \includegraphics[scale=0.4]{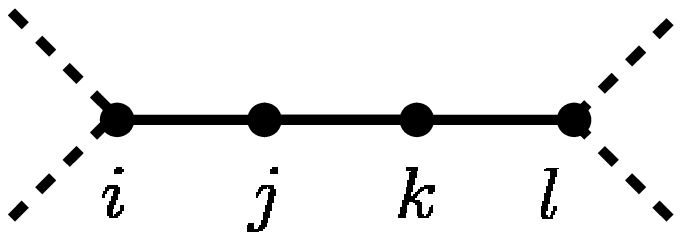}  
\caption{Node $j$ is degree 2.}
 \label{fig4}
\end{minipage}
\hspace{-2.5cm}
\begin{minipage}{.65\linewidth}
\hspace{3.2cm}
\includegraphics[scale=0.4]{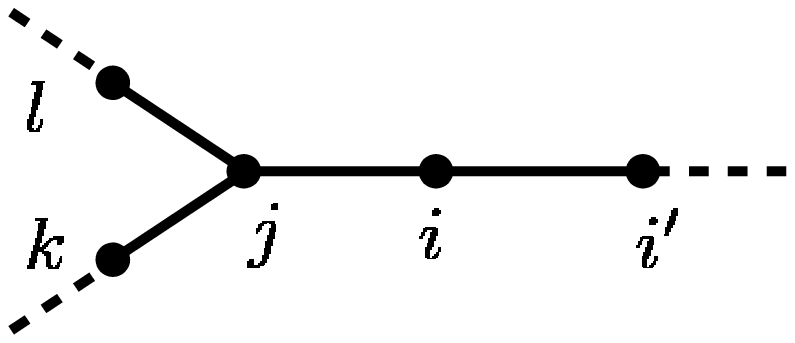}  
\caption{Node $j$ is degree 3, node $i$ is degree 2.}
 \label{fig5}
\end{minipage}
\end{figure}

\begin{figure}
\hspace{0.5cm}
\includegraphics[scale=0.6]{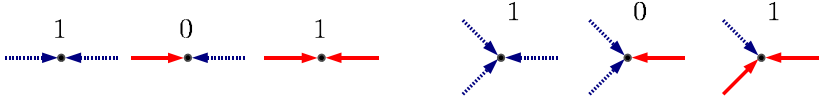}
\vspace{-0.6cm}
\caption{Rules at degree 2 and 3 nodes.}
 \label{fig6}
\end{figure}

By splitting nodes as in \fref{fig3},
any graphs can be transformed so that every node is of degree at most three. 
We make the compatibility functions of new edges infinitely strong:
$\psi_{ij}(x_i,x_j)=\delta_{x_i,x_j}$.
This change does not affect the fixed points of LBP algorithm, hence
the Bethe approximation.
In subsections \ref{4.1} and \ref{4.2} we assume that the graph $G$ has
undergone this transformation.
The same transformation also appears in \cite{f1966} and \cite{ccPlanar}.
The first messages $\mu$ are defined on this transformed graph.
We define ``secondary messages'' $\{\nu_{(i,j)}\}$ 
on the transformed graph using $\{\mu_{(i,j)}\}$.
First, let $j$ be a node of degree two such that at least one adjacent node,
denoted by $k$,
has degree two (\fref{fig4}).  
We define $\nu$ at the node $j$ by the following conditions:
\begin{eqnarray}
 \sum_{x_j} \nu_{(j,i)}(x_j) \mu_{(j,k)}(x_j) = 0 , \quad
 \sum_{x_j} \mu_{(j,i)}(x_j) \nu_{(j,k)}(x_j) = 0   \label{deg2orth} ,\\
 \sum_{x_j} \nu_{(j,i)}(x_j) \nu_{(j,k)}(x_j) = 1 \label{deg2normal}.
\end{eqnarray}
Similarly, 
let $j$ be a node of degree two such that at least one adjacent node,
denoted by $i$, has degree two (\fref{fig5}).  
We define $\nu$ at node $j$ by the following conditions:
\begin{eqnarray}
\fl
 \sum_{x_j} \nu_{(j,i)}(x_j) \mu_{(j,k)}(x_j) \mu_{(j,l)}(x_j) = 0   ,\quad 
 \sum_{x_j} \mu_{(j,i)}(x_j) \nu_{(j,k)}(x_j) \mu_{(j,l)}(x_j) = 0  \nonumber
, \\
\fl
 \sum_{x_j} \mu_{(j,i)}(x_j) \mu_{(j,k)}(x_j) \nu_{(j,l)}(x_j) = 0  \label{deg3orth} 
, \\ 
\fl
 \sum_{x_j} \nu_{(j,i)}(x_j) \nu_{(j,k)}(x_j) \mu_{(j,l)}(x_l) = 1   , \quad
 \sum_{x_j} \mu_{(j,i)}(x_j) \nu_{(j,k)}(x_j) \nu_{(j,l)}(x_j) = 1     \nonumber
 , \\
\fl
 \sum_{x_j} \nu_{(j,i)}(x_j) \mu_{(j,k)}(x_j) \nu_{(j,l)}(x_j) = 1   . \label{deg3normal}
\end{eqnarray}
These conditions determine $\nu_{(j,i)}$ uniquely up to a scalar factor
at nodes of degree two, and up to sign at nodes of degree three.
We assume that the first component $\nu_{(i,j)}(+)$ is negative without
loss of generality.
The above relations and $\sum_{x_j}\prod_{i\in N(j)}\mu_{(j,i)}(x_j)=1$ \eref{equal1}
are pictorially summarized in \fref{fig6}.
These cases are sufficient because we can add a node of degree two at any
edge if necessary.
We call such diagrams {\it propagation diagrams}.
The blue dashed and the light red arrows express $\mu$ and $\nu$,
respectively.

A condition which is similar to (\ref{deg2orth}) and (\ref{deg3orth}) is 
imposed in \cite{Loopqary} to deduce the update rule of the LBP algorithm,
though they consider a graphical model with variables on the edges.

\subsection{Propagation rules} \label{4.2}
\begin{figure}
\begin{center}
\includegraphics[scale=0.5]{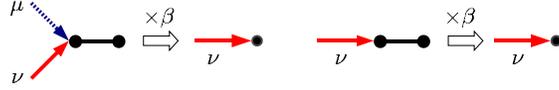}   
\caption{Propagation rules.}
\label{fig7}
\end{center}
\end{figure}

The rules depicted in \fref{fig6} describe what happens when messages collide
at a node.
The next two lemmas show how first and secondary messages propagate.
\begin{lem} \label{lem1}
See \fref{fig4}. Suppose nodes $j$ and $k$ are of degree 2. 
Then, there is $\beta_{jk} \in \mathbb{R}$ such that
\begin{eqnarray}
\fl
 \beta_{jk} \nu_{(j,k)}(x_j) = \sum_{x_k} \psi_{jk}(x_j,x_k)
 \nu_{(k,l)}(x_k)   , \qquad
 \beta_{jk} \nu_{(k,j)}(x_k) = \sum_{x_j} \psi_{kj}(x_k,x_j)
 \nu_{(j,i)}(x_j) .  \label{2deg2}
\end{eqnarray}
\end{lem}
\begin{proof}
Using (\ref{BPupdatec}) and (\ref{deg2orth}),
$ \sum_{x_j,x_k} \mu_{(j,i)}(x_j) \psi_{(j,k)}(x_j,x_k) \nu_{(k,l)}(x_k) = 
 \sum_{x_k} \mu_{(k,j)}(x_k) \nu_{(k,l)}(x_k) =0$.
Hence $\sum_{x_k}\psi_{jk}(x_j,x_k) \nu_{(k,l)}(x_k) \propto \nu_{(j,k)}(x_j)$. 
The proportion is equal in the both directions $(j,k)$ and $(k,j)$ because of
 the condition (\ref{deg2normal}).
\end{proof}
Next we proceed to a degree three node.
\begin{lem} \label{lem2}
See \fref{fig5}. Suppose node $j$ is of degree three and $i$ is of degree two.
Then, there is $\beta_{ji} \in \mathbb{R}$ such that
\begin{eqnarray}
 \beta_{ji} \nu_{(j,i)}(x_j) = \sum_{x_i} \psi_{ji}(x_j,x_i) 
 \nu_{(i,{i}')}(x_i)  ,\\
 \beta_{ji} \nu_{(i,j)}(x_i) = \sum_{x_j} \psi_{ji}(x_j,x_i) 
 \nu_{(j,l)}(x_j) \mu_{(j,k)}(x_j), \label{lemma2eq2}\\
\beta_{ji} \nu_{(i,j)}(x_i) = \sum_{x_j
} \psi_{ji}(x_j,x_i) 
 \mu_{(j,l)}(x_j) \nu_{(j,k)}(x_j). \label{lemma2eq3}
\end{eqnarray}
\end{lem}
\begin{proof}
 We can show in the same way as the previous lemma.
\end{proof}
These lemmas say that the secondary messages $\nu$ propagate 
with rate $\beta$ for both directions, though 
the first messages $\mu$ propagate without variation in scales.
Equations \eref{lemma2eq2} and \eref{lemma2eq3}
hold when the adjacent nodes $i$ and $j$ have degree three
as in \fref{fig8}. 
We associate numbers $\beta_{ij}$ for all undirected edges in $E$.
Propagation diagrams of these results are summarized in \fref{fig7}. 

\subsubsection{Two more rules}
\begin{figure}
\begin{center}
\includegraphics[scale=0.4]{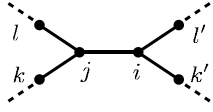}  
\caption{Node $j$ and $i$ have degree three.}
\label{fig8}
\end{center}
\end{figure}

In addition to the rules in \fref{fig6},
we have to give a rule for the case in
which three secondary messages $\nu$ collide at a node.
\begin{lem}\label{lemgamma}
Let $N(j)=\{i,k,l\}$, then
\begin{eqnarray}
\fl
\quad
\sum_{x_i}\nu_{(j,i)}(x_j)\nu_{(j,k)}(x_j)\nu_{(j,l)}(x_j)
=
\frac{1}{\sqrt{b_j(+)b_j(-)}}(b_j(+)-b_j(-))
=: \gamma_{j} . \label{gamma}
\end{eqnarray}
\end{lem}
\begin{proof}
We need direct computation of the message vectors.
By the orthogonal condition \eref{deg3orth},
 $\nu_{(j,i)}$, $\nu_{(j,k)}$ and $\nu_{(j,l)}$ are determined up to
 scalar factors.
The scales are determined by \eref{deg3normal}.
Since we assume the first components of $\nu$ are negative, $\nu$ must have
 the following form:
\begin{eqnarray}
\fl
\nu_{(j,i)}(x_j)=
-x_j
\sqrt{
\frac{\mu_{(j,i)}(1)\mu_{(j,i)}(-1)}
{\mu_{(j,k)}(1)\mu_{(j,k)}(-1)\mu_{(j,l)}(1)\mu_{(j,l)}(-1)} 
}
\mu_{(j,k)}(-x_j)\mu_{(j,l)}(-x_j). \label{nuexpression}
\end{eqnarray}
From \eref{1marginalc} the result follows.
\end{proof}
We use the next lemma when we split a node as in \fref{fig3}
\begin{lem}\label{lembeta}
See \fref{fig8}. If the nodes $j$ and $i$ are of degree three and 
$\psi_{i,j}{(x_i,x_j)}=\delta_{x_i,x_j}$, 
 then $\beta_{ij}=1$.  
\end{lem}
\begin{proof}
By \eref{lemma2eq3}, \eref{nuexpression}
and the remark after lemma \ref{lem2},
\begin{eqnarray}
 \beta_{ij}\nu_{(i,j)}(x_i)
&=\sum_{x_j}\psi_{ij}(x_i,x_j)\mu_{(j,l)}(x_j)\nu_{(j,k)}(x_j)
\nonumber \\
&=\mu_{(j,l)}(x_i)\nu_{(j,k)}(x_i) \nonumber \\
&=-x_i
\sqrt{
\frac{\mu_{(j,k)}(1)\mu_{(j,k)}(-1)\mu_{(j,l)}(1)\mu_{(j,l)}(-1)}
{\mu_{(j,i)}(1)\mu_{(j,i)}(-1)} 
}\mu_{(j,i)}(-x_i). \nonumber
\end{eqnarray}
Since $\mu_{(j,i)}(x_i)=\mu_{(i,l')}(x_i)\mu_{(i,k')}(x_i)$ and
$\mu_{(i,j)}(x_i)=\mu_{(j,l)}(x_i)\mu_{(j,k)}(x_i)$, we can show
 $\beta_{ij}=1$ by using \eref{nuexpression} for $(i,j)$.
\end{proof}

\subsection{At general nodes} \label{4.3}
\begin{figure}[h]
\begin{center}
 \includegraphics[scale=0.5]{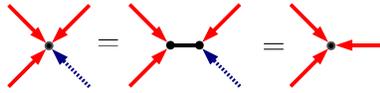} 
\caption{Reduction to the case of $n=d_j$.}
\label{fig9}
\end{center}
\end{figure}

\begin{figure}[h]
\begin{center}
 \includegraphics[scale=0.5]{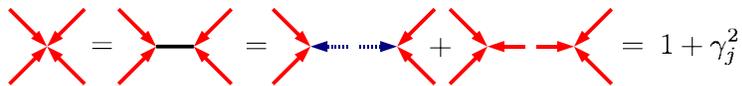} 
\caption{Derivation of the inductive relation in the case of $n=4$.}
\label{fig10}
\end{center}
\end{figure}

We have defined $\nu$ on the transformed graph $G_T=(V_T,E_T)$ in which 
the degree of every node is at most three. 
By shrinking added edges, we define $\nu$ on the original graph $G=(V,E)$. 
In other words the injection $\vec{E} \rightarrow \vec{{E}_T}$ induces
the messages on the original graph from the transformed graph.
At each node of the original graph, we show that the following theorem holds.
This theorem generalizes the rules in \fref{fig6} and lemma \ref{lemgamma}.
\begin{thm} \label{thm2}
Let $j \in V$ and $N(j)=\{i_{1},\ldots,i_{d_j}\}$. Then, we have
\begin{eqnarray}
f_{n}(\gamma_{j})=\sum_{x_j}\prod_{s=1}^n 
\nu_{(j,i_s)}(x_{j}) \prod_{t=1}^{d_j-n} \mu_{(j,i_{n+t})}(x_{j}) ,
\end{eqnarray}
where $\{f_n(x)\}_{n=0}^{\infty}$ is a set of polynomials
defined inductively by the relations $f_0(x)=1,f_1(x)=0$ and $f_{n+1}(x)=x f_n(x) + f_{n-1}(x)$.
\end{thm}
\begin{proof}
We can reduce to the case of $n=d_j$ by splitting node $j$ and using
the propagation rules in \fref{fig7}.
See \fref{fig9} for example.
The proof is done by induction;
the cases of $n=1,2,3$ is obtained by definition.
In the case of $n$ we split the node $j$ into $j'$ and $j''$ where $2$
and $n-2$ secondary messages join at $j'$ and $j''$, respectively.
Adding a node $k$ between $j'$ and $j''$, we use the relation
$\mu_{(k,j')}(x_k)\mu_{(k,j'')}({x_k}')+\nu_{(k,j')}(x_k)\nu_{(k,j'')}({x_k}')=
 \delta_{x_k,{x_k}'}$.
By lemma \ref{lembeta} and the induction hypothesis, we obtain
$f_n(\gamma_j)=f_{2}(\gamma_{j}) f_{n-2}(\gamma_{j})+f_{3}(\gamma_j) f_{n-1}(\gamma_j)$.
The assertion follows from the definition of the polynomials $f_n$.
\Fref{fig10} illustrate this procedure.
\end{proof}
It is surprising that the right hand side is determined by the 
value $\gamma_j$ which depends only on the belief $b_j$, while the messages
$\mu_{(j,i_s)}$ and $\nu_{(j,i_s)}$ are not determined by $b_j$.

In this section we have derived a set of rules.
In addition to the diagrams to show the rules,
diagrams in which these rules are successively applied are also 
called propagation diagrams.
%
\subsection{In the case of one dimensional systems} 
In the case of one dimensional spin systems, results of section 3 and 4 are
reduced to easy observations.
Let the graph $G$ be a cycle of length $N$. 
We cut the node $1$ as discussed in section 3.1 to obtain a string.
The partition function is 
\begin{eqnarray}
 Z= \sum_{ x_1= \pm 1} T^{x_1}_{\quad x_{\bar{1}}}
\end{eqnarray}
where the transfer matrix $T=T_1$ is defined by (\ref{t}) and (\ref{ts}).
Therefore $Z$ is equal to the sum of the first and second eigenvalues of the
matrix $T$.
By theorem \ref{BPeqThm}, the first messages $\mu_1$ and $\mu_{\bar{1}}$
are the first right and left eigenvectors of the transfer matrix,
where the first eigenvalue is $Z_B$.
By lemma \ref{lem1}, we see that $\nu_1$ and $\nu_{\bar{1}}$ are the second
right and left eigenvectors, and the second eigenvalue is $Z_B \prod \beta_{ji}$.
The conditions of (\ref{deg2orth}) are regarded as orthogonality of
eigenvectors. 
The results of these sections are a generalization to the transfer tensor
associated with more complicated graphs with nodes of degree three.

\section{Loop series expansion formula} 
\subsection{Derivation of loop series expansion formula} \label{5.1}
Let $G=(V,E)$ be a graph, and $D$ be a subset of the edge-set $E$.
An edge-induced subgraph of $D$ is a subgraph of the graph $G$ whose
edge-set is $D$ and whose node-set consists of all nodes that are
incident with at least one edge in $D$. 

We are now ready to prove the expansion formula of the partition
function.
\begin{thm} \label{thmloop}
For each solution of (\ref{BPeq1}),(\ref{BPeq2}), define
 $\{\beta_{ij}\},\{\gamma_i\}$ by the above lemmas. Then the following 
expansion formula holds.
\begin{eqnarray}
Z=Z_B(1+\sum_{\phi \neq C \in \mathcal{G}} r(C)) ,\label{loopexpansion} \\
r(C):=\prod_{ij \in {E}_C}\beta_{ij} 
\prod_{i \in V_C} f_{d_{i}(C)}(\gamma_{i}), \nonumber
\end{eqnarray}
Where $\mathcal{G}$ is the set of all edge-induced subgraphs of $G$
and $C=(V_C,E_C)$.
\end{thm}
\begin{proof}
Adding a node $k$ on each edge $(i,j)$ of $G$, we expand the partition
 function by the relation
$\mu_{(k,i)}(x_k)\mu_{(k,j)}({x_k}')+\nu_{(k,i)}(x_k)\nu_{(k,j)}({x_k}')=
 \delta_{x_k,{x_k}'}$.
We have $2^{|E|}$ terms in this expansion, because for each edge $(i,j)$ there
 is a choice: 
$\mu_{(k,i)}(x_k)\mu_{(k,j)}({x_k}')$ or $\nu_{(k,i)}(x_k)\nu_{(k,j)}({x_k}')$.
If we regard the $\nu$-edges as a edge-set, each term can be identified with an edge-induced graph.
By theorem \ref{thm2} and lemma \ref{lem1} and \ref{lem2}, the term corresponding to $C$ is
 equal to $r(C)$.
\end{proof}
Since $f_{1}(x)=0$, $C$ makes a contribution to the sum only if $C$ does
not have a node of degree one in $C$. Such $C$ is called a generalized loop. 
In the case that $G$ is a tree, there is no generalized loop,
therefore $Z=Z_B$. This is well-known \cite{Pearl}. 

Note that $f_n(0)=1$ if $n$ is even, and $f_n(0)=0$ if $n$ is odd.
If $\gamma_i=0$ for all $i \in V$, only generalized loops with even
degrees contribute to the sum.
This is reminiscent of the high temperature expansion of the Ising model
without magnetic field.
We further discuss this point in section \ref{isingregular}.
In addition, if the graph is planar, 
these terms are summed by a single Pfaffian 
\cite{f1966,ccPlanar}.

We can choose and expand 
some of the edges step by step  
with propagation diagrams as in
\ref{appexample}, while we expand at all edges
in the above proof.

Let
$\Theta(\{\beta_{ij}\},\{\gamma_i\}):=1+\sum_{\phi \neq C \in
\mathcal{G}} r(C)$ . 
This is a polynomial of indeterminates $\{\beta_{ij}\}$ and
$\{\gamma_i\}$, and its coefficients are positive
integers, because $f_n(x)$ are polynomials with coefficients positive
integers.
We can assign other quantities to nodes and edges such as $\{m_i\}$ and
$\{\tau_{ij}\}$ discussed later in section \ref{5.2}, and
represent the formula in different manner
with these quantities, but such assignment may cause
large positive and negative coefficients.
Advantage of using $\gamma$ and $\beta$ is that the coefficients are not huge
because they are all positive and the total sum is determined by
$L$ as discussed later
in \ref{SumofCoefficients}.
Note that the method of propagation diagrams gives an algorithm for computing $\Theta$.
%

The messages $\mu$ and $\nu$ are explicitly given if we change the
compatibility functions.
By the definitions of the beliefs \eref{1marginal} and \eref{2marginal}, we see that
\begin{eqnarray}
 \prod_{(i,j)\in E} \psi_{ij}(x_i,x_j)  \propto \prod_{(i,j)\in E} b_{ij}(x_i,x_j) 
\prod_{i \in V} \frac{1}{{b_{i}(x_i)}^{d_i-1}}.
\end{eqnarray}
Therefore, we can retake compatibility functions as
\begin{eqnarray}
 \psi_{ij}(x_i,x_j)=\frac{b_{ij}(x_i,x_j)}{{b_i(x_i)}^{(d_i-1)/d_i}{b_j(x_j)}^{(d_j-1)/d_j}} \label{repara}
\end{eqnarray}
without changing the joint probability distribution.
Moreover, this does not cause any change to the result of the LBP
algorithm, namely beliefs.
In the rest of section \ref{5.1} and section \ref{5.2}, we assume this
representation of compatibility functions. 
By \eref{BPupdatec}, for all $(j,i)\in \vec{E}$ the first messages 
have the following forms:
$\mu_{(j,i)}(x_j)= b_{j}(x_j)^{1/{d_j}}$.
The secondary messages are determined by theorem \ref{thm2} as
\begin{eqnarray}
 \nu_{(j,i)}(x_j)=\frac{-x_j
  b_j(-x_j)^{(d_j-1)/d_j}}{b_j(+)^{(d_j-2)/2d_j}b_j(-)^{(d_j-2)/2d_j}}
\quad {}^{\forall}(j,i)\in \vec{E}. \label{reparanu}
\end{eqnarray}
Using lemma \ref{lem1} and lemma \ref{lem2} on the transformed graph,
we see that 
\begin{eqnarray}
\fl
 \beta_{ij}=\sum_{x_i,x_j}
\frac{b_{ij}(x_i,x_j)}{{b_i(x_i)}^{(d_i-1)/d_i}{b_j(x_j)}^{(d_j-1)/d_j}}
\nu_{(i,i_1)}(x_i)\nu_{(j,j_1)}(x_j)
\prod_{s=2}^{d_i-1}\mu_{(i,i_s)}(x_i)
\prod_{t=2}^{d_j-1}\mu_{(j,j_t)}(x_j), 
\end{eqnarray}
where $N(i)=\{j,i_1,\ldots,i_{d_{i}-1}\}$ and $N(j)=\{i,j_1,\ldots,j_{d_j-1}\}$.
A direct computation shows that,
\begin{eqnarray}
 \beta_{ij}=\frac{b_{ij}(+,+)b_{ij}(-,-)-b_{ij}(+,-)b_{ij}(-,+)}
{\sqrt{b_i(+)b_i(-)}\sqrt{b_j(+)b_j(-)}}
\quad
{}^{\forall}(i,j)\in {E}
. \label{beta}
\end{eqnarray}
\Eref{beta} implies that $|\beta_{ij}| \leq 1$.
By (\ref{2marginal}),
$\beta_{ij}=0$ if and only if $\psi_{ij}$ can be factorized by some
functions as
$\psi_{ij}(x_i,x_j)=\psi_{i}(x_i)\psi_{j}(x_j)$; no interaction
between node $i$ and $j$.

\subsection{Relation to the result of Chertkov and Chernyak} \label{5.2}
Chertkov and Chernyak \cite{Loop} show the loop series expansion formula for
general vertex models and factor graph models, which are more general
than pairwise interaction models considered in this paper.
Focusing on pairwise interaction models, however, we found
the further relations of the first and secondary messages 
in theorem \ref{thm2}, which derives
our representation of $r(C)$.
In this section we show that the expansion formula given in theorem
\ref{thmloop} is equivalent to the result of Chertkov and 
Chernyak in \cite{Loop}.
Let us briefly review their result in the case of pairwise MRF:
\begin{eqnarray}
 Z=Z_B(1+\sum_{C \in \mathcal{G}}\check{r}(C)), \quad
\check{r}(C)=\prod_{ij \in E_{C}} \tau_{ij}\prod_{i \in V_{C}}\rho_{i}(C) ,\\
\rho_{i}(C)=\frac{(1-m_i)^{d_i(C)-1}+(-1)^{d_i(C)}(1+m_i)^{d_i(C)-1}}
{2(1-m_i^2)^{d_i(C)-1}}  , \nonumber \\
\tau_{ij}=\sum_{x_{i},x_j}b_{ij}(x_i,x_j)(x_i-m_i)(x_j-m_j), \quad
 m_i=b_i(1)-b_i(-1). \nonumber
\end{eqnarray}
It suffices to prove $r(C)=\check{r}(C)$ for all generalized loops $C$.

By the inductive definition of the polynomials $f_n$, we see that
\begin{eqnarray}
f_n(x)=\frac{\lambda_1^{n-1}-\lambda_2^{n-1}}{\lambda_1-\lambda_2},
\end{eqnarray}
where
$\lambda_1,\lambda_2$ are the roots of the quadratic equation
$\lambda^2-x\lambda -1=0$.
Using the definition of $\gamma_i$, direct calculation derives
\begin{eqnarray}
\rho_{i}(C)(2\sqrt{b_i(1)b_i(-1)})^{d_i(C)}=f_{d_i(C)}(\gamma_i). \label{eq1}
\end{eqnarray}
Using (\ref{beta}), we see that
\begin{eqnarray}
 \tau_{ij}=\beta_{ij}(2 \sqrt{b_i(1)b_i(-1)}) (2 \sqrt{b_j(1)b_j(-1)})  .\label{eq2}
\end{eqnarray}
Combining (\ref{eq1}) and (\ref{eq2}) gives the claim.

We append two comments on the difference between our approach and theirs.
First, we derived the expansion from the viewpoint of message
passing operation.
The quantities $\beta_{ij}$ and $\gamma_{i}$ are characterized by
propagation of messages.
On the other hand, Chertkov and Chernyak used covariances and means of
the beliefs for the expansion. 
Secondly, we interpreted the recursion relation of $f_n$ by transformations of
the graphs in the proof of theorem \ref{thm2},
though the corresponding relation is not clear
in their choice of variables $\tau_{ij}$ and $\rho_i$.
The recursion is effectively used for upper bounding the number of
generalized loops in section 5.4.2.
%
\subsection{Ising partition function on a regular graph} 
\label{isingregular}
In this section we briefly discuss the connection between the polynomial
$\theta(\beta,\gamma):=\Theta(\{\beta_{ij}=\beta\},\{\gamma_i=\gamma\})$ 
and the partition function of the Ising model on a regular
graph $G$. 
A graph $G$ is called regular if all of the degrees of nodes are the same.
We see in corollary \ref{p} that $\theta$ can be regarded as a transform of the partition
function on the basis of the Bethe approximation,
and apply it to the derivation of susceptibility formula.
In this subsection, we assume
$\gamma_i=\gamma$, $\beta_{ij}=\beta$
and $G$ is regular graph of degree $d$.

Since we have relations  (\ref{gamma}) and (\ref{beta}),
we solve $b_{ij}$ by $\beta$ and $\gamma$ as
\begin{eqnarray}
\fl
 b_{ij}(x_i,x_j)=
\frac{1}{4}
\left(
1+
\frac{x_i \gamma}{\sqrt{4+\gamma^2}}+
\frac{x_j \gamma}{\sqrt{4+\gamma^2}}+
\frac{x_ix_j \gamma^2}{{4+\gamma^2}}
\right) 
+
\frac{x_ix_j \beta_{}}{{4+\gamma^2}} . \label{bij}
\end{eqnarray}
By (\ref{loopexpansion}) and (\ref{repara}), 
the polynomial $\theta$ admits the following identity:
\begin{eqnarray}
\theta(\beta,\gamma)
=
\sum_{x_i}
\prod_{(i,j)} b_{ij}(x_i,x_j)
\prod_{i} b_i(x_i)^{1-d} \label{thetaidentity}
\end{eqnarray}
where $b_{ij}$ is defined by (\ref{bij}) and $b_{i}$ is defined by
$\sum_{x_j=\pm1}b_{ij}$. 
Let $Z(K,h)=\sum_{x_i}\exp({K\sum_{ij}x_ix_j+h\sum_{i}x_i})$
be a partition function.
As a function of $y$ and $z$, we obtain the following identity.

\begin{cor} \label{p}
For a regular graph $G$ of degree $d$,
 \begin{eqnarray}
\fl
  \theta(\beta,\gamma)=Z(K,h) 
\Big(\frac{\sqrt{1-z^2}(1+y^2z)}{1-y^2z^2}\Big)^{|E|}
\Big(\frac{\sqrt{(1-y^2)(1-y^2z^2)}}{2(1+y^2z)}\Big)^{|V|}
 \end{eqnarray}
where $\beta={(1-y^2)z}/{(1-y^2z^2)}$ and
 $\gamma={2y(1+z)}/{\sqrt{(1-y^2)(1-y^2z^2)}}$.
Furthermore,  $K=\tanh^{-1}z $ and
\begin{eqnarray}
\exp({2h})=
\Big(\frac{1+y}{1-y}\Big)
\Big(\frac{1+yz}{1-yz} \Big)^{1-d} .
\end{eqnarray}
\end{cor}
\begin{proof}
The proof is accomplished by calculations based on (\ref{thetaidentity}).
Let $b_{ij}(x_i,x_j)=\exp({Kx_ix_j+h'x_i+h'x_j+C})$ and 
$b_{i}(x_i)=\exp({h''x_i+D})$.
We define $z=\tanh K$ and $y=\tanh h'$.
By (\ref{bij}), $\beta$, $\gamma$ and $C$ are solved by $y$ and $z$.
The condition $b_{i}=\sum_{x_j=\pm1}b_{ij}$ determines $h''$ and $D$.
Let $h=dh'+(1-d)h''$, then the product of $b_{ij}$ and $b_{i}$ is
 proportional to $\exp({K\sum_{ij}x_ix_j+h\sum_{i}x_i})$.
\end{proof}

If $y=0$, then  $h=0$, $\gamma=0$ and $\beta=z$. 
This theorem is reduced to the well known high temperature expansion. 
Therefore this formula is an extension of the high temperature expansion of
the Ising model with external field.

We proceed to obtain a formula of zero field susceptibility which is
defined by
\begin{eqnarray}
 \chi(K):=\frac{d}{dh}{\log Z(K,h)}\Big|_{h=0}.
\end{eqnarray}
By the differentiation of (\ref{thetaidentity}), we have 
\begin{eqnarray}
\fl
(1+z-dz)^2 \chi(K) =
(1+z)(1+z-dz)
+&
2z(z^2-1) 
\frac{\partial}{\partial z} \log {\theta(z,0)} \nonumber
\\ 
&+ 
8(1+z)^2 
\frac{\partial}{\partial \gamma^2} \log \theta(z,\gamma)\Big|_{\gamma=0}.
\end{eqnarray}
If we substitute $1$ for $\theta$, which corresponds to the Bethe approximation, 
this formula reduces to the well known formula of Bethe
approximation of susceptibility \cite{ds1957}.
Higher order approximation can be obtained by enumerating generalized
loops that appear in $\theta$.
Comparison of traditional ways of enumeration of subgraphs 
 \cite{ds1957crystal,s1961} and our
expansion is an interesting future research topic.
\subsection{Miscellaneous topics}
\subsubsection{Representation of $T_k$}
Using the first and secondary messages and $\{\beta_{ij}\}$, 
${T_k}^{x_k}_{\quad {x_{\bar{k}}}}$ defined in \eref{ts}
admits a simple representation.
Let $k$ be a leaf node of the tree $\hat{G}$ and
$N(k)=\{i,j\}$ on the original graph $G$,
and let $i_0,i_1,\ldots,i_l$ be the unique path from 
$i_0=k$ to $i_l=\bar{k}$ on $\hat{G}$, where $i_1=j,i_{l-1}=i$. 
It is easy to see that $T_k\mu_{(k,i)}=\mu_{(k,i)}$ and 
$T_k\nu_{(k,i)}=\prod_{t=1}^{l}\beta_{i_{t-1}i_t}\nu_{(k,i)}$ with
propagation diagrams on $\hat{G}$.
Therefore
\begin{eqnarray}
 {T_k}^{x_k}_{\quad {x_{\bar{k}}}}= \mu_{(k,i)}(x_{\bar{k}})\mu_{(k,j)}(x_{k})+
\prod_{t=1}^{l}\beta_{i_{t-1}i_t}\nu_{(k,i)}(x_{\bar{k}})\nu_{(k,j)}(x_{k}).
\end{eqnarray}
This shows that $\nu_{(k,j)}$ and $\nu_{(k,i)}$ are the left and right
eigenvectors of the matrix $T_k$ and their eigenvalue is
$\prod_{t=1}^{l}\beta_{i_{t-1}i_t}$.

\subsubsection{The number of generalized loops}\label{SumofCoefficients}
\begin{figure}
\includegraphics[scale=0.4]{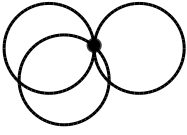}
\vspace{-0.3cm}
\caption{$L=3$ case.}
 \label{fig11}
\end{figure}

We first show that the polynomial $\theta(1,\gamma)$ depends on the
graph $G$ only through $L$, the number of linearly independent cycles.
Since $\beta=1$, we can shrink any edges without changing corresponding
polynomial $\theta$. 
Any graph with $L$ independent cycles can be reduced to a
graph in which only one node has degree more than two. 
See \fref{fig11}: $L$ rings are joined at one point. 
Therefore, cutting $L$ loops, we obtain
\begin{eqnarray}
\theta(1,\gamma)&=\sum_{k=0}^{L} {L \choose k}f_{2k}(\gamma) \nonumber \\
&=\Biggl(\frac{2\sqrt{4+\gamma^2}}{\gamma+\sqrt{4+\gamma^2}}\Biggr)^{L-1}
+
\Biggl(\frac{2\sqrt{4+\gamma^2}}{-\gamma+\sqrt{4+\gamma^2}}\Biggr)^{L-1} .\label{eq5}
\end{eqnarray}
This equality shows that the sum of coefficients of $\Theta$ is equal to
\begin{eqnarray}
 \theta(1,1)= \Biggl(\frac{5-\sqrt{5}}{2}\Biggr)^{L-1}
+
\Biggl(\frac{5+\sqrt{5}}{2}\Biggr)^{L-1}.
\end{eqnarray}
Moreover, we obtain a bound for the number of generalized loops.
\begin{cor}
Let $\mathcal{G}_{0}$ be the set of all generalized loops of $G$ including
 empty set. Then,
\begin{eqnarray}
\left| \mathcal{G}_{0} \right|
\leq
 \Biggl(\frac{5-\sqrt{5}}{2}\Biggr)^{L-1}
+
\Biggl(\frac{5+\sqrt{5}}{2}\Biggr)^{L-1}.
\end{eqnarray}
This bound is attained if and only if every node of
a generalized loop has the degree at most three.
\end{cor}
\begin{proof}
Since $f_n(1)>1$ for all $n>4$ and $f_2(1)=f_3(1)=1$,
we have $r(C)|_{\beta=\gamma=1} \geq 1$ for all $C \in \mathcal{G}_{0}$
, and the equality holds if and only if 
$d_i(C) \leq 3$
for all $i\in V_{C}$.
This shows $\left| \mathcal{G}_{0} \right| \leq \theta(1,1)$ and
the equality condition.
\end{proof}
%
%
\section{Marginal expansion formula}
\subsection{Derivation of the marginal expansion formula}
\begin{figure}
\includegraphics[scale=0.7]{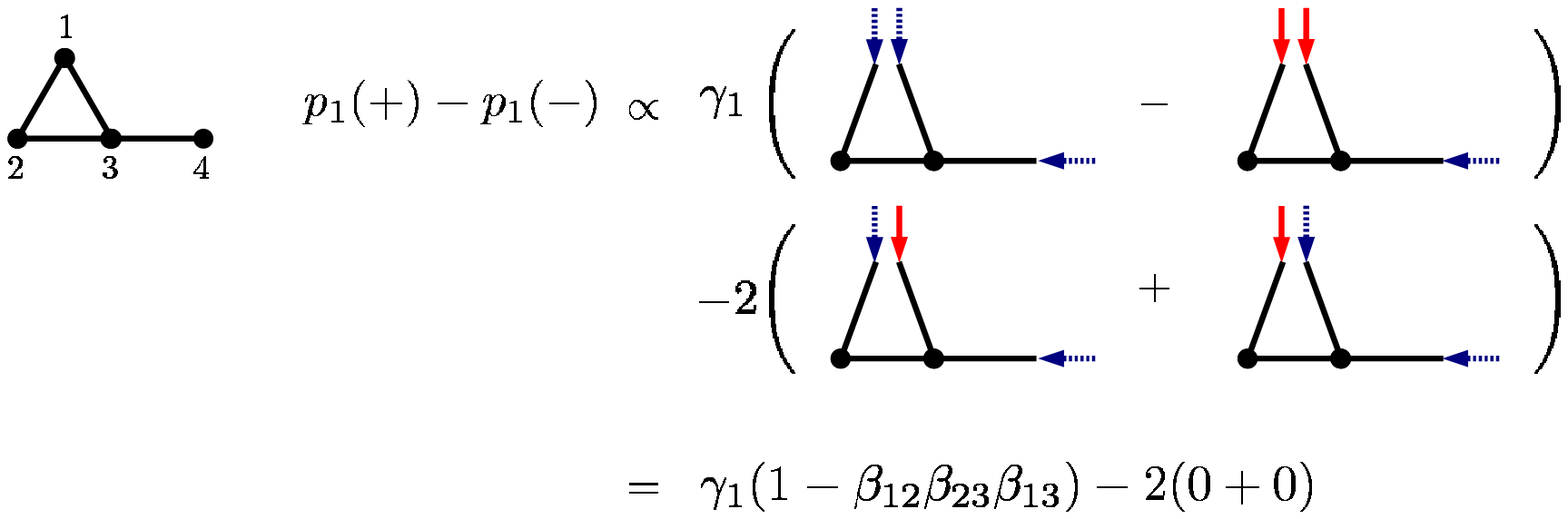}
\vspace{-1.2cm}
\caption{An example of expansion with propagation diagrams in one loop cases.}
 \label{fig12}
\end{figure}

We show relations between the approximated marginal $b_k$
and the true marginal $p_k$. 
In this section, we take compatibility functions as \eref{repara} once again. 
Without loss of generality, we assume $k=1$ and
$N(1)=\{i,j\}$, i.e. node $1$ has degree two.
Indeed, split of nodes as in \fref{fig3} gives the same marginal
probability to the added nodes as the original one.
Because
\begin{eqnarray}
 \mu_{(1,i)}=\mu_{(1,j)}={\sqrt{b_1(+)} \choose \sqrt{b_1(-)}},\qquad 
 \nu_{(1,i)}=\nu_{(1,j)}={-\sqrt{b_1(-)} \choose \sqrt{b_1(+)}},  \nonumber
\end{eqnarray}
we see the following equations by a direct computation:
\begin{eqnarray}
\fl
\left[
\begin{array}{cc}
1 & 0 \\
0 & 0 
\end{array}
\right]
= b_1(+)\mu_{(1,i)} \mu_{(1,j)}^T +b_1(-)\nu_{(1,i)} \nu_{(1,j)}^T
\nonumber \\
-\sqrt{b_1(+)b_1(-)}\mu_{(1,i)}\nu_{(1,j)}^T
-\sqrt{b_1(+)b_1(-)}\nu_{(1,i)}\mu_{(1,j)}^T , \label{identity1}\\
\fl
\left[
\begin{array}{cc}
0 & 0 \\
0 & 1 
\end{array}
\right]
= b_1(-)\mu_{(1,i)} \mu_{(1,j)}^T +b_1(+)\nu_{(1,i)} \nu_{(1,j)}^T
\nonumber \\
+\sqrt{b_1(+)b_1(-)}\mu_{(1,i)}\nu_{(1,j)}^T
+\sqrt{b_1(+)b_1(-)}\nu_{(1,i)}\mu_{(1,j)}^T . \label{identity2}
\end{eqnarray}
By the definition of marginal probability, we see that
\begin{eqnarray}
  \frac{Z}{Z_B} p_1(\pm) = \sum_{x_1,x_{\bar{1}}} 
{T_1'}^{x_1}_{\quad {x_{\bar{1}}}} I_{\{x_1=x_{\bar{1}}=\pm\}},
\end{eqnarray}
where ${T_1'}^{x_1}_{\quad {x_{\bar{1}}}}:=
\sum_{x_2,\ldots,x_L}
T^{x_1,\ldots,x_L}_{\quad x_{{1}},\ldots,x_{{L}}}$ and $I$ is the
indicator function.
Using these equations, we can show the following theorem.
%
\begin{thm} \label{thmMarginalExpansion}
\begin{eqnarray}
\fl
  \frac{Z}{Z_B}p_{1}(\pm) =
b_{1}(\pm)\left(\sum_{x_1}{T_1'}^{x_1}_{\quad
	 {x_{\bar{1}}}}\mu_{(1,i)}(x_1)\mu_{(1,j)}(x_1)\right)
+b_{1}(\mp)\left(\sum_{x_1}{T_1'}^{x_1}_{\quad
	 {x_{\bar{1}}}}\nu_{(1,i)}(x_1)\nu_{(1,j)}(x_1)\right)
\nonumber \\
\fl
\mp \sqrt{b_1(+)b_1(-)}
\left(\sum_{x_1}{T_1'}^{x_1}_{\quad{x_{\bar{1}}}}\mu_{(1,i)}(x_1)\nu_{(1,j)}(x_1)
+\sum_{x_1}{T_1'}^{x_1}_{\quad{x_{\bar{1}}}}\nu_{(1,i)}(x_1)\mu_{(1,j)}(x_1)   \label{MarginalExpansionFormula}
\right).
\end{eqnarray}
The four summation terms appeared in (\ref{MarginalExpansionFormula}) can be
 expanded with propagation diagrams.
\end{thm}
This expansion is 
computationally intractable if $L$ is large, in a similar way to
the partition function $Z$.
For relatively small graphs, however,
we may be able to expand these terms.
The terms in the expansion of the first two summations are labelled by the
generalized loops, while the other terms in expansion of the last
two summations are labelled by other subgraphs: each subgraph does not
have nodes of degree one except the node $1$.
The expansion may be heuristically used for
approximate computation of marginal probability distributions, 
namely we can correct beliefs using terms corresponding to major
subgraphs in expanded representation of
(\ref{MarginalExpansionFormula}).
With this theorem, an already known fact is easily deduced \cite{1loop}.
\begin{cor}\label{cor1loop}
 Letting $L$=1 and node $1$ is on the unique cycle in $G$, 
 $p_1(+)-p_1(-)$ and $b_{1}(+)-b_{1}(-)$ have the same sign.
\end{cor}
\begin{proof}
By theorem \ref{thmMarginalExpansion},
\begin{eqnarray}
\fl
\frac{Z}{Z_B}\frac{p_{1}(+)-p_{1}(-)}{\sqrt{b_1(+)b_1(-)}}
=\gamma_{1}
\Biggl(
\sum_{x_1}{T_1'}^{x_1}_{\quad
{x_{\bar{1}}}}\mu_{(1,i)}(x_1)\mu_{(1,j)}(x_1)
-\sum_{x_1}{T_1'}^{x_1}_{\quad
{x_{\bar{1}}}}\nu_{(1,i)}(x_1)\nu_{(1,j)}(x_1)
\Biggr) \nonumber \\
-2
\Biggl(\sum_{x_1}{T_1'}^{x_1}_{\quad{x_{\bar{1}}}}\mu_{(1,i)}(x_1)\nu_{(1,j)}(x_1)
+\sum_{x_1}{T_1'}^{x_1}_{\quad{x_{\bar{1}}}}\nu_{(1,i)}(x_1)\mu_{(1,j)}(x_1)   
\Biggr).
\end{eqnarray}
The right hand side can be expanded with propagation diagrams as in \fref{fig12}.
The first summation is 1, the second is a product of $\beta$, and
 the third and fourth is equal to 0.
Since $\left|\beta_{ij}\right| \leq 1$ the result follows. 
\end{proof}
A problem of finding an assignment that maximize the marginal
probability $p_1$ is called maximum marginal assignment problem \cite{1loop}.
This corollary asserts that the assignment that maximize the belief
$b_1$ is the solution of this problem.
%
%
%
\subsection{Example}
Consider a graph in \fref{fig2}. In this case
theorem \ref{thmMarginalExpansion}
turns out to be 
\begin{eqnarray}
\fl
 \frac{Z}{Z_B}p_{1}(\pm)=b_{1}(\pm)\Biggl(1+\beta_{23}\beta_{34}\beta_{24}\Biggr)
+b_{1}(\mp)\beta_{14}\beta_{13}\Biggl(\beta_{34}+\beta_{34}\beta_{24}\beta_{23}\gamma_{3}\gamma_{4}\Biggr)
\nonumber \\
\mp \sqrt{b_1(+)b_1(-)}\Biggl(\beta_{13}\beta_{34}\beta_{23}\beta_{24}\gamma_{3}
+\beta_{14}\beta_{34}\beta_{23}\beta_{24}\gamma_{4}\Biggr) .\label{exampleMarginalExpansion}
\end{eqnarray}
By (\ref{exampleMarginalExpansion}), we have
\begin{eqnarray}
\fl
 \frac{Z}{Z_B}\frac{p_{1}(+)-p_{1}(-)}{\sqrt{b_1(+)b_1(-)}}
=\gamma_{1}\Biggl(
1+\beta_{23}\beta_{34}\beta_{24}-\beta_{14}\beta_{13}\beta_{34}
-\beta_{14}\beta_{13}\beta_{34}\beta_{24}\beta_{23}\gamma_{3}\gamma_{4}
\Biggr) \nonumber \\
-2\Biggl(\beta_{13}\beta_{34}\beta_{23}\beta_{24}\gamma_{3}
+\beta_{14}\beta_{34}\beta_{23}\beta_{24}\gamma_{4}\Biggr) .\label{exampleMAP}
\end{eqnarray}
The right hand side discriminates which state is more plausible.
Let $\beta_{34}=0$ in (\ref{exampleMAP}), this expression is reduced to
the case of $L=1$ and consistent with the result of corollary \ref{cor1loop}.
Let $\beta_{13}=0$. In this case we see that $p_{1}(+)-p_{1}(-)$ does not necessarily
have the same sign as $\gamma_{1}$ \cite{1loop}.

If, for example, $|\beta_{ij}|\leq 1/2$, 
$|\gamma_3|,|\gamma_4|\leq 1$ and 
$|\gamma_3|/2,|\gamma_2|/2 \leq |\gamma_1|$,
then we see from \eref{exampleMAP} that
$p_{1}(+)-p_{1}(-)$ and $\gamma_1$ have the same sign.
The first condition requires weakness of the interactions, and the second
condition requires that the beliefs at the nodes of degree three are not too
much biased.
The last condition is satisfied if $p_{1}(+)$ and $p_{1}(-)$ are not too close to
each other.

If we take variables $m_i$ and $\tau_{ij}$ in section \ref{5.2}
instead of $\gamma_i$ and
$\beta_{ij}$, the expressions (\ref{exampleMarginalExpansion}) and
(\ref{exampleMAP}) become more complicated in general, and it is hard
to find simple conditions. 
\section{Concluding remarks}
We introduced propagation diagrams that enable us to compute loop
series expansion of a partition function and marginal distributions with
a set of simple rules. In this method, parameters $\beta_{ij}$ and
$\gamma_{i}$ are naturally assigned to each edge and node.

Accuracy of the Bethe approximation depends both on the strength of interactions and
the topology of the underlying graph. 
The effect of the interactions is captured by the values of $\beta$ and $\gamma$. 
The topological aspect of the graph, in
the sense of Bethe approximation, is extracted in the polynomial $\Theta$.

We suggest future research topics.
First, understanding of the structure of the polynomial $\Theta$
is important to construct efficient
approximation algorithms exploiting graph topology.
The properties of $\Theta$ should be investigated further.
Secondly, on the basis of the results of this paper,
it is interesting to understand the  
empirically known fact: if LBP does not converge, the quality of
the Bethe approximation is low \cite{mk2005}.
Since we show a direct relation between the message passing operation and 
the expansion variables $\beta$ and $\gamma$, convergence of the LBP
algorithm can be analyzed using them.


\ack
This work was supported in part by Grant-in-Aid for JSPS Fellows
20-993 and Grant-in-Aid for Scientific Research (C) 19500249.

\section{References}
\bibliographystyle{iopart-num}

\appendix
\section*{Appendix}
\setcounter{section}{1}  

\subsection{Example of expansion}\label{appexample}
We consider the graph in \fref{fig2}.
Normalizing $Z_B=1$, we can calculate the loop expansion of the
partition function as in the following figure, using propagation diagrams.
 \begin{figure}[h]
\begin{center}
\includegraphics[scale=0.5]{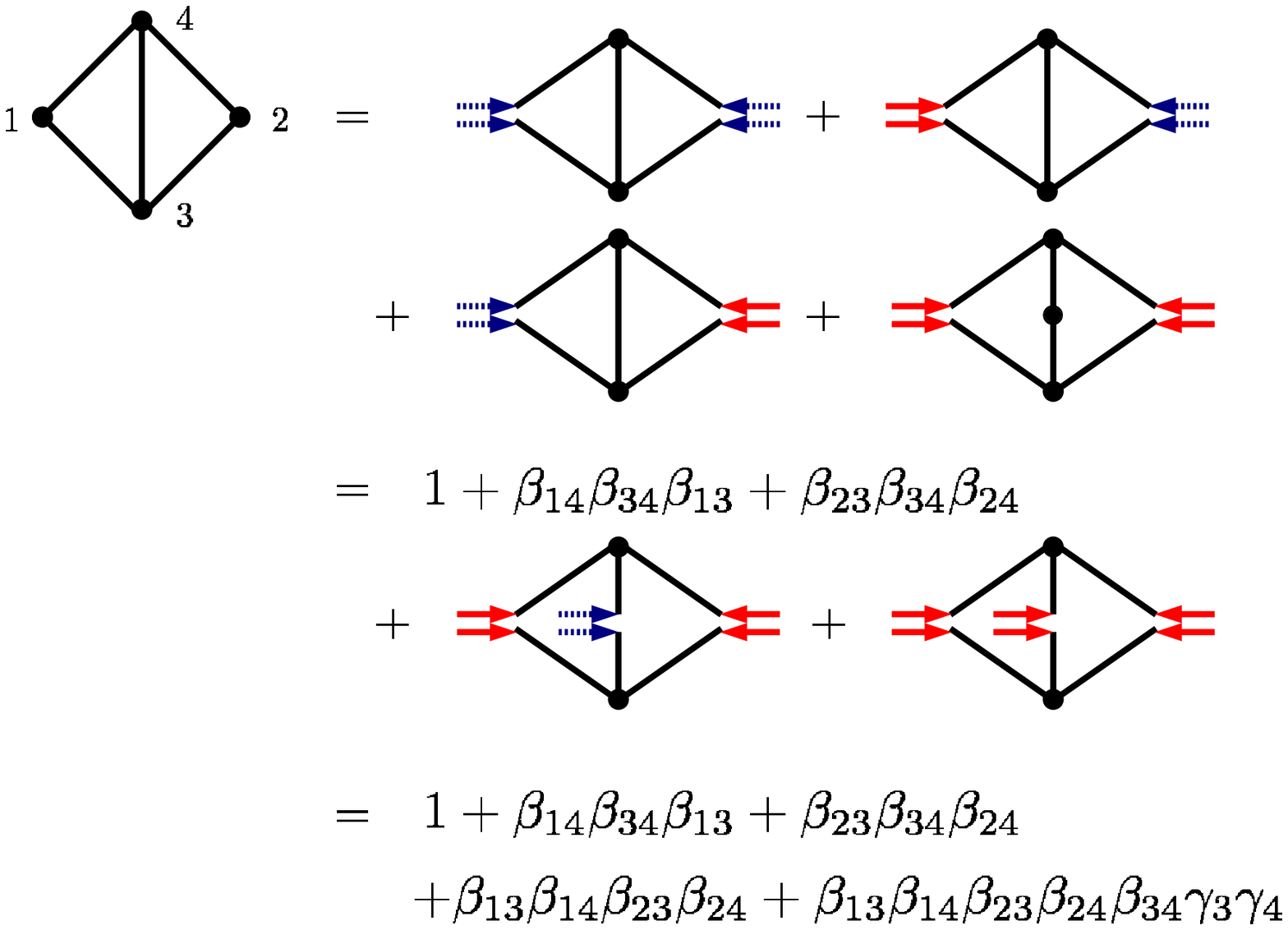}
\vspace{-0.4cm}
\caption{Expansion of the partition function with the method of propagation
  diagrams.
It is not necessary to expand all edges as in the proof of theorem
  \ref{thmloop}. }
 \label{figA1}
\end{center}
  \end{figure}

\begin{figure}[h]
\begin{center}
\includegraphics[scale=0.5]{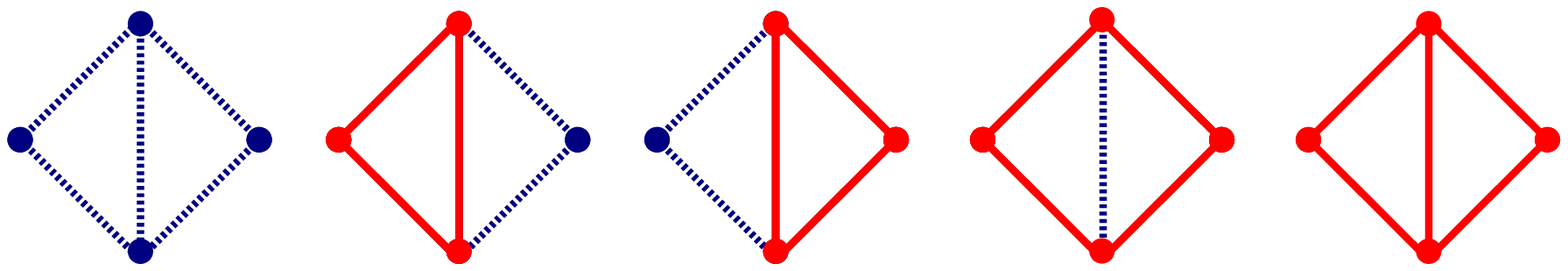}
\vspace{-0.4cm}
\caption{Light red parts are generalized loops.}
 \label{figA2}
\end{center}
  \end{figure}

Five terms in the final expression of \fref{figA1} correspond to the subgraphs in 
\fref{figA2} respectively.

\end{document}